\documentclass[journal ]{new-aiaa}
\usepackage[utf8]{inputenc}
\usepackage{textcomp}

\usepackage{graphicx}
\usepackage{amsmath}
\usepackage[version=4]{mhchem}
\usepackage{siunitx}
\usepackage{longtable,tabularx}
\usepackage{bbm}
\usepackage{algorithm}
\usepackage{algpseudocode}
\usepackage{multirow}
\title{Analysis and Design of Spare Strategy \\ for Large-Scale Satellite Constellation 
\\ Using Direct Insertion under $(r,q)$ policy}

\author{Seungyeop Han \footnote{Ph.D. Candidate, Daniel Guggenheim School of Aerospace Engineering}}
\author{Zachary Grieser \footnote{Ph.D. Student, Daniel Guggenheim School of Aerospace Engineering}}
\affil{Georgia Institute of Technology, Atlanta, Georgia, 30332}
\author{Shoji Yoshikawa\footnote{Chief Expert, Advanced Technology R\&D Center, Mitsubishi Electric Corporation}}
\author{Takumi Noro\footnote{Researcher, Advanced Technology R\&D Center, Mitsubishi Electric Corporation}}
\author{Takumi Suda\footnote{Head Researcher, Advanced Technology R\&D Center, Mitsubishi Electric Corporation}}
\affil{Mitsubishi Electric Corporation, Amagasaki 661-8861, Japan}
\author{Koki Ho \footnote{Dutton-Ducoffe Professor, Associate Professor, Daniel Guggenheim School of Aerospace Engineering, AIAA Senior Member, kokiho@gatech.edu (Corresponding Author)}}
\affil{Georgia Institute of Technology, Atlanta, Georgia, 30332}

\begin{document}
\maketitle

\footnotetext{This paper is a substantially revised version of a portion of Paper AAS 24-164 presented at 2024 AAS/AIAA Astrodynamics Specialist Conference, Broomfield, CO, August 11-15, 2024.}

\begin{abstract}
This paper introduces a Markov chain–based approach for the analysis and optimization of spare-management policies in large-scale satellite constellations. Focusing on the direct strategy, we model spare replenishment as a periodic-review reorder-point/order-quantity policy, where spares are deployed directly to constellation planes. The stochastic behavior of satellite failures and launch vehicle lead times is captured through Markov representations of both failure and replenishment dynamics. Based on this efficient and accurate framework, we construct and solve an optimization problem aimed at minimizing operational costs. The effectiveness of the proposed method is demonstrated through a case study using a real-world mega-constellation.
\end{abstract}

\section*{Nomenclature}
{\renewcommand\arraystretch{1.0}
\noindent\begin{longtable*}{@{}l @{\quad=\quad} l@{}}
$\tau_\text{mc}$ & Time step of the discrete-time Markov process, in days \\
$\lambda_\text{sat}$ & Failure rate of a satellite, in failures per unit time \\
$\mu_\text{lv}$ & Mean interval between launches, in days \\
$\tau_\text{lv}$ & Constant launch order processing time, in days \\
$q$ & Replenishment quantity for in-plane spares \\
$r$ & Reorder point for in-plane spares \\
$ N_{\text{sat}} $ & 
Maximum in-plane state level including operational and spare satellites \\
$\bar{N}_\text{sat}$ & Nominal number of operational satellites per in-plane orbit \\
$ N_{\text{orbit}}$  & Number of in-plane (constellation) orbital planes \\
$ P_f$ & Failure transition matrix for in-plane \\
$ P_q$ & $q$-unit replenishment transition matrix for in-plane/parking states\\
$\pi^{q}$ & Expected in-plane state distribution immediately after $q$-unit replenishment \\
$\pi^{r}$ & Expected in-plane state distribution at the $r$-reorder point \\
$\pi^{\text{io}}$ & Expected in-plane state distribution during the inter-order (IO) period \\
$\pi^{\text{lt}}$ & Expected in-plane state distribution during the lead-time (LT) period \\
$\pi^{\text{rc}}$ & Expected in-plane state distribution over the full replenishment cycle (RC) \\
\end{longtable*}}

\section{Introduction}
\lettrine{L}{arge}-scale satellite constellations require effective spare-management policies to maintain performance in the presence of satellite failures. For mega-constellations in low Earth orbit (LEO), it is widely believed that launching replacement satellites is more cost-effective than on-orbit repair. This is due to declining launch costs, mass-production efficiencies, and the high expense of designing satellites to be serviceable. These trends motivate the need to study spare-management strategies tailored for such systems.

Two main approaches have been proposed for spare management in LEO constellations: direct and indirect replenishment strategies. The indirect strategy employs a large launch vehicle (LV) to deliver batches of spare satellites into parking orbits, from which they are later transferred to the constellation planes once alignment is achieved through RAAN drift caused by J2 perturbation. This method benefits from batch discounts and lower per-unit launch costs but suffers from longer replenishment delays due to the slow orbital drift. In contrast, the direct strategy uses a small LV to deliver spare satellites directly to the constellation’s in-plane orbits, enabling immediate replenishment at the expense of higher launch costs per satellite.

This paper revisits the direct spare strategy, extending the work in~\cite{han2024analysis}, and provides detailed analysis of the counterpart policy used for comparison in our recent study~\cite{han2025indirect}. It contains supplementary material not explicitly addressed in~\cite{han2025indirect}, focusing on the modeling and evaluation of the direct strategy. For a detailed literature survey and discussion of the indirect strategy, readers are referred to~\cite{han2025indirect}.

The remainder of the paper is organized as follows. Section~\ref{sec2} introduces the modeling preliminaries. Section~\ref{sec3} presents the analytical method for evaluating the direct resupply strategy. Section~\ref{sec4} applies this method to assess system performance. Section~\ref{sec5} validates the model through Monte Carlo simulation, and Section~\ref{sec6} demonstrates its application in a design optimization context. Finally, Section~\ref{sec7} concludes the paper.

\section{Preliminaries} \label{sec2}
\subsection{Spare Management Policy}
\subsubsection{Direct Resupply Strategies}
\begin{figure}[!h]
    \centering
    \includegraphics[width=.45\textwidth]{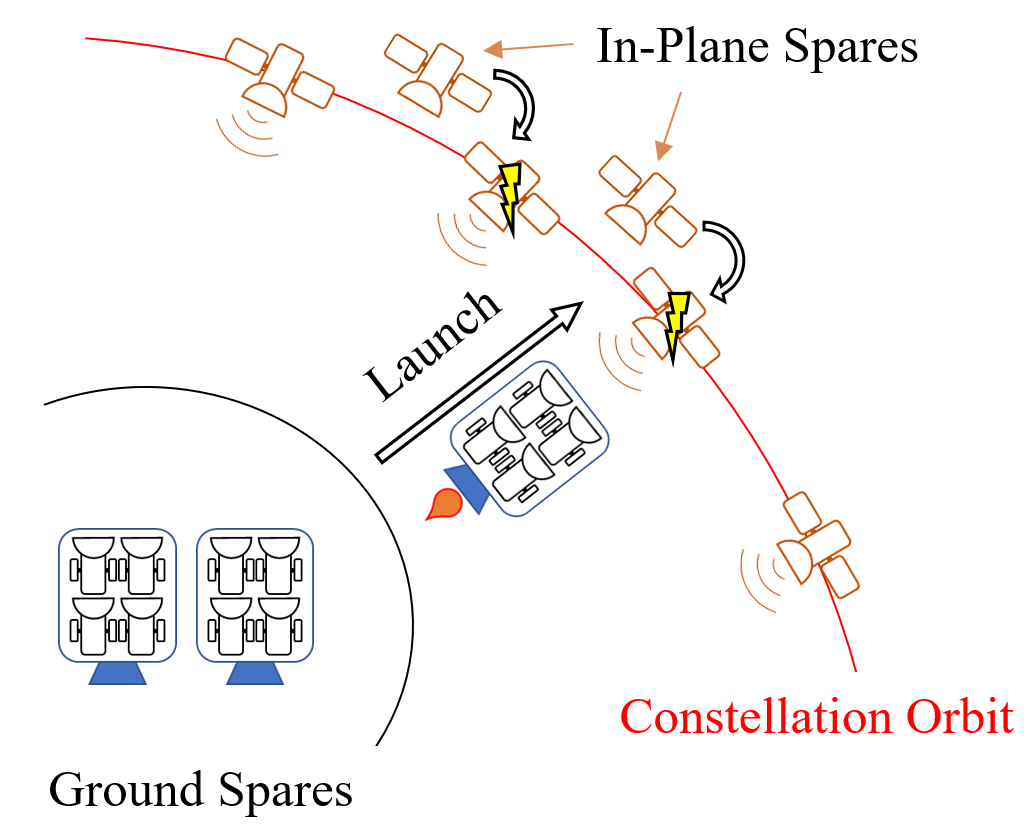}
    \caption{Illustration of Direct Spare Strategy}
    \label{fig_strategy}
\end{figure}

The direct strategy uses a small LV to deliver spares to an in-plane orbit when replenishment is needed. Figure~\ref{fig_strategy} illustrates the direct strategy. If a failure occurs, an in-plane spare immediately replaces the failed satellite, and whenever the number of in-plane spares falls below a threshold, a ground resupply order is placed, and the replacement arrives after the LV’s lead time.

\subsubsection{Inventory Management Policy} \label{sec:inventory_model} 
The spare management of the direct strategy is modeled using an $(r,q)$ policy, which works as follows: if the stock level is less than or equal to $r$, an order of size $q$ is placed; otherwise, no order is made. Each order arrives after its lead time, with no additional orders placed between review points.

\begin{figure}[!ht]
    \centering
    \includegraphics[width=.45\textwidth]{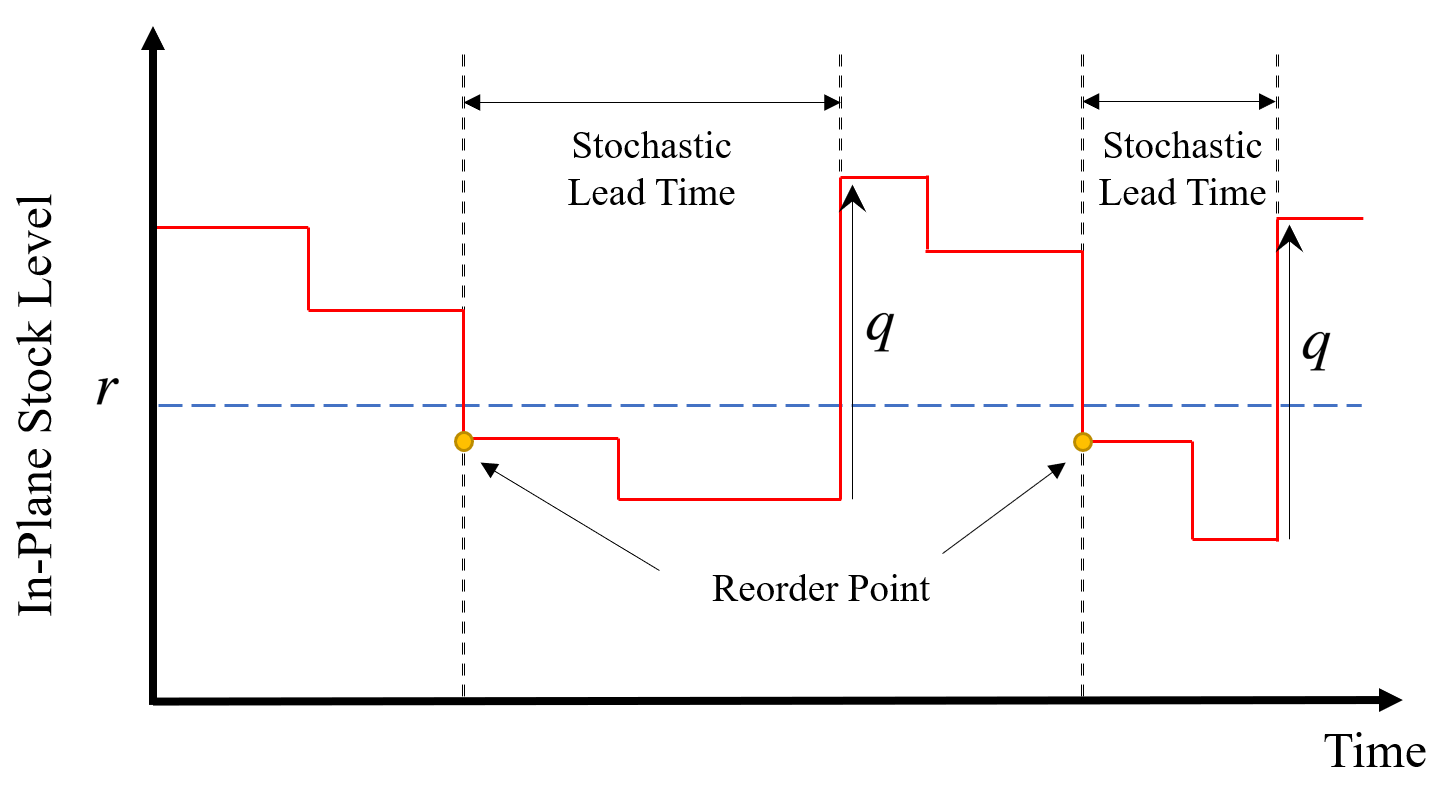}
    \caption{Stock level profile of constellation orbits under $(r, q)$ policy}
    \label{fig:inventory_model}
\end{figure}

\subsection{Constellation Model}
This research focuses on large scale constellations in LEO, specifically those with the same number of satellites in each orbital plane. Unlike the indirect strategy, the direct strategy does not require a symmetrical RAAN distribution or identical orbital inclinations across planes. In this configuration, the constellation consists of $N_{\text{orbit}}$ in-plane orbital planes, each nominally populated with $\bar{N}_{\text{sat}}$ satellites.

\subsection{Markov Chain Model}
We model the spare‐satellite count as a discrete‐time Markov chain.  Let \(X_k\in\{0,1,\dots,N_{\rm sat}\}\) be the number of satellites (including spares) at step \(k\), and write the distribution
\begin{equation} 
    \pi_k = \begin{bmatrix}
        \mathbb{P}(X_k = N_\text{sat}) & \mathbb{P}(X_k = N_\text{sat}-1) & \cdots & \mathbb{P}(X_k = 0)
    \end{bmatrix}^\top,\quad \pi_k(i)=\mathbb{P}(X_k=i)
\end{equation}
Assuming a time‐homogeneous transition matrix $P\in\mathbb R^{(N_{\rm sat}+1)\times(N_{\rm sat}+1)}$ with entries $P_{ij}=\mathbb{P}(X_{k+1}=i\mid X_k=j)$, the chain evolves by $\pi_{k+1}=P\,\pi_k$. Under the usual ergodicity conditions (e.g.\ $P$ irreducible and aperiodic), the Markov Chain has a unique $\pi$ satisfying
\begin{equation} 
    \pi = P \pi
\end{equation}
and it gives the long‐run fraction of time the chain spends in each state. In our spare‐management model, failures (state decreases) and replenishment (state increases) guarantee these conditions.

\subsection{Probabilistic Model}
The probabilistic modeling framework used in this section, including the satellite failure distribution, lead-time modeling, and transition matrix construction, is directly adopted from our previous work~\cite{han2025indirect}. 

\subsubsection{Satellite Failure Probability Distribution }
Let $\tau_\text{mc}$ denote the time step of the Markov process, and let $\lambda_\text{sat}$ be the failure rate of an operational satellite per $\tau_\text{mc}$. Then, the probability of observing $k$ failures from $n$ satellites (including spares) during $\tau_\text{mc}$ is given by:
\begin{equation}\label{mpois_fail}
    \nu_{k,n} = 
    \mathbb{P}(F=k| X=n) = \begin{cases}
        0 & \mbox{if } k > \bar{N}_\text{sat}\\ 
        \frac{(n\lambda_\text{sat})^k}{k!}e^{-n\lambda_\text{sat}} & \mbox{if } n \leq \bar{N}_\text{sat} \mbox{ and } k \leq \bar{N}_\text{sat} \\ 
        \frac{(\bar{N}_\text{sat}\lambda_\text{sat})^k}{k!}e^{-\bar{N}_\text{sat}\lambda_\text{sat}} & \mbox{if } n > \bar{N}_\text{sat} \mbox{ and } k \leq \bar{N}_\text{sat} \\
    \end{cases}
\end{equation}
where $F$ is the number of failures, $\bar{N}_\text{sat}$ is the nominal number of operational (non-spare) satellites. This formulation assumes immediate failure replacement and that spare satellites do not fail (i.e., $k>\bar{N}_\text{sat}$). When $n \leq \bar{N}_\text{sat}$, all satellites are operational, yielding a total failure rate of $n\lambda_\text{sat}$. When $n > \bar{N}_\text{sat}$, the number of operational satellites is fixed at $ \bar{N}_\text{sat}$, making a total failure rate of $\bar{N}_\text{sat} \lambda_\text{sat}$, and excess satellites are treated as spares.

Additionally, let $N_{\text{sat}_\text{c}}$ be the maximum number of satellites including spares in constellation orbits. Then the state transition matrix due to failure can be defined as:
\begin{equation} \label{eq:fail_matrix_sim}
    P_{f} = \begin{bmatrix}
    v_{0,N_{\text{sat}_\text{c}}} 
      & 0                     & \cdots         & 0         \\
    v_{1,N_{\text{sat}_\text{c}}} 
      & v_{0,N_{\text{sat}_\text{c}}-1}   & \cdots         & 0         \\
    \vdots 
      & \vdots                & \ddots & \vdots       \\
    1-\sum_{k=0}^{N_{\text{sat}_\text{c}}} v_{k,N_{\text{sat}_\text{c}}}
      & 1-\sum_{k=0}^{N_{\text{sat}_\text{c}}-1} v_{k,N_{\text{sat}_\text{c}}-1} 
                             & \cdots 
                                       & v_{0,0}  
\end{bmatrix}
\end{equation}
where $P_f \in \mathbb{R}^{( N_{\text{sat}}+1 )\times ( N_{\text{sat}}+1)}$. By construction, each column vector sums to one, and the matrix is lower triangular, clearly showing that multiplying by $P_f$ always decrease the state level. In summary, if $\pi$ is the in-plane state distribution, then $P_f\pi$ gives the distribution after a one-step failure.

\subsubsection{LV Lead‐Time Probability Distribution}
The ground‐resupply lead time is modeled as a shifted exponential distribution \cite{jakob2019optimal}: $T \sim \text{Exp}(\mu_\text{lv}) + \tau_\text{lv}$, and its probability density function is
\begin{equation}
    f(T=t;\mu_\text{lv},\tau_\text{lv}) = 
    \begin{cases}
        \frac{1}{\mu_\text{lv}} e^{-{(t-\tau_\text{lv})}/{\mu_\text{lv}}} & \quad t \geq \tau_\text{lv}\\
        0 & \quad t < \tau_\text{lv}\\
    \end{cases}
\end{equation}
where $\mu_\text{lv}$ is the mean of the exponential component and $\tau_\text{lv}$ is the  fixed LV‐processing delay. To simplify our discrete‐time modeling, we choose $\tau_\text{mc}$ such that $\tau_\text{lv}$ is an integer multiple of $\tau_\text{mc}$. Then the probability of having a lead time between $k$ and $k+1$ time steps of $\tau_\text{mc}$ is computed as
\begin{equation} \label{eq:rho_lv}
\begin{aligned}
    \rho_{k+1} &= \mathbb{P}(k\tau_\text{mc} \leq T < (k+1) \tau_\text{mc}) \\
    &= 
    \begin{cases}
        e^{-k\tau_\text{mc}/{\mu_\text{lv}}}\left( 1 - e^{-\tau_\text{mc}/{\mu_\text{lv}}}\right), & \text{if } k\tau_\text{mc} \geq \tau_\text{lv} \\
        0, & \text{otherwise } \\
    \end{cases}
\end{aligned}
\end{equation}
Note that each orbit is assumed to place at most one LV order at a time.

\subsubsection{State Space and Reorder Threshold Projections}
The maximum number of satellites in a constellation orbit is \(N_{\text{sat}} = q + r\), so the state distribution \(\pi^{(\cdot)}\) lies in \(\mathbb{R}^{N_{\text{sat}} + 1}\). To apply the \((r, q)\) policy, we need to isolate the portion of \(\pi\) corresponding to states where the stock level is less than or equal to \(r\). For this purpose, we define the following projection matrices:
\begin{equation} \label{eq:Cr_matrix}
    C_{r}^+ = \begin{bmatrix}
        I_{N_{\text{sat}}-r} & \textbf{0}_{(N_{\text{sat}}-r)\times (r+1)} \\ \textbf{0}_{(r+1) \times (N_{\text{sat}}-r)} & \textbf{0}_{r+1}
    \end{bmatrix},\ 
    C_{r}^- = \begin{bmatrix}
        \textbf{0}_{N_{\text{sat}}-r} & \textbf{0}_{(N_{\text{sat}}-r)\times (r+1)} \\ \textbf{0}_{(r+1) \times (N_{\text{sat}}-r)} & I_{r+1}
    \end{bmatrix}
\end{equation}
Then $C_{r}^+\pi$ gives the distribution for $X>r$ and $C_{r}^-\pi$ the distribution for $X\leq r$.  These projections are key for deriving the failure and replenishment transition matrix.

\subsubsection{Replenishment Transition Matrix}
After the lead time elapses, the system receives \(q\) spare satellites. To model the corresponding state update, we define the replenishment transition matrix \(P_q\), which maps the distribution immediately before delivery to the distribution immediately after:
\begin{equation} \label{eq:resupply_trans_matrix}
    P_{q} = 
    \begin{bmatrix}
    \begin{array}{c|c}
        I_q & I_{r+1} \\
        \mathbf{0}_{(r+1)\times q} & \mathbf{0}_{q\times (r+1)}
    \end{array}
\end{bmatrix}
\end{equation}
and $P_{q} \in \mathbb{R}^{( N_{\text{sat}}+1) \times ( N_{\text{sat}}+1)}$. In summary, $P_{q}\pi$ gives the distribution after receiving $q$ spares, when $\pi$ was the distribution immediately before replenishment.

\section{Modeling and Analysis of Spare Management Policy} \label{sec3}
In this section, we present the analysis method of the in-plane $(r,q)$-policy introduced in Section \ref{sec:inventory_model} for the direct resupply, utilizing the Markov chain.
The approach uses $\pi^q$ and $\pi^r$, which represent the conditional state probability distributions (i.e., the distributions at specific events), to formulate a repeated Markov chain. The overall procedure as follows. First, determine the relationship between $\pi^q$ and $\pi^r$ and compute the stationary solution of them. Next, $\pi^\text{io}$ and $\pi^\text{lt}$ are computed using $\pi^q$ and $\pi^r$, respectively. Finally, $\pi^\text{rc}$ is expressed as a linear combination of $\pi^\text{io}$ and $\pi^\text{lt}$. 

When we discretize the time step, we assume that delivery occurs first, followed by failure, and finally the reorder is made. The dynamic modeling is based on this assumption. While this may deviate from continuous-time behavior, the ambiguity introduced by discretization becomes negligible as $\tau_\text{mc}$ approaches zero.

\subsection{Transition Equation from Delivery to Reorder} 
\begin{figure}[!ht]
    \centering
    \includegraphics[width=.55\textwidth]{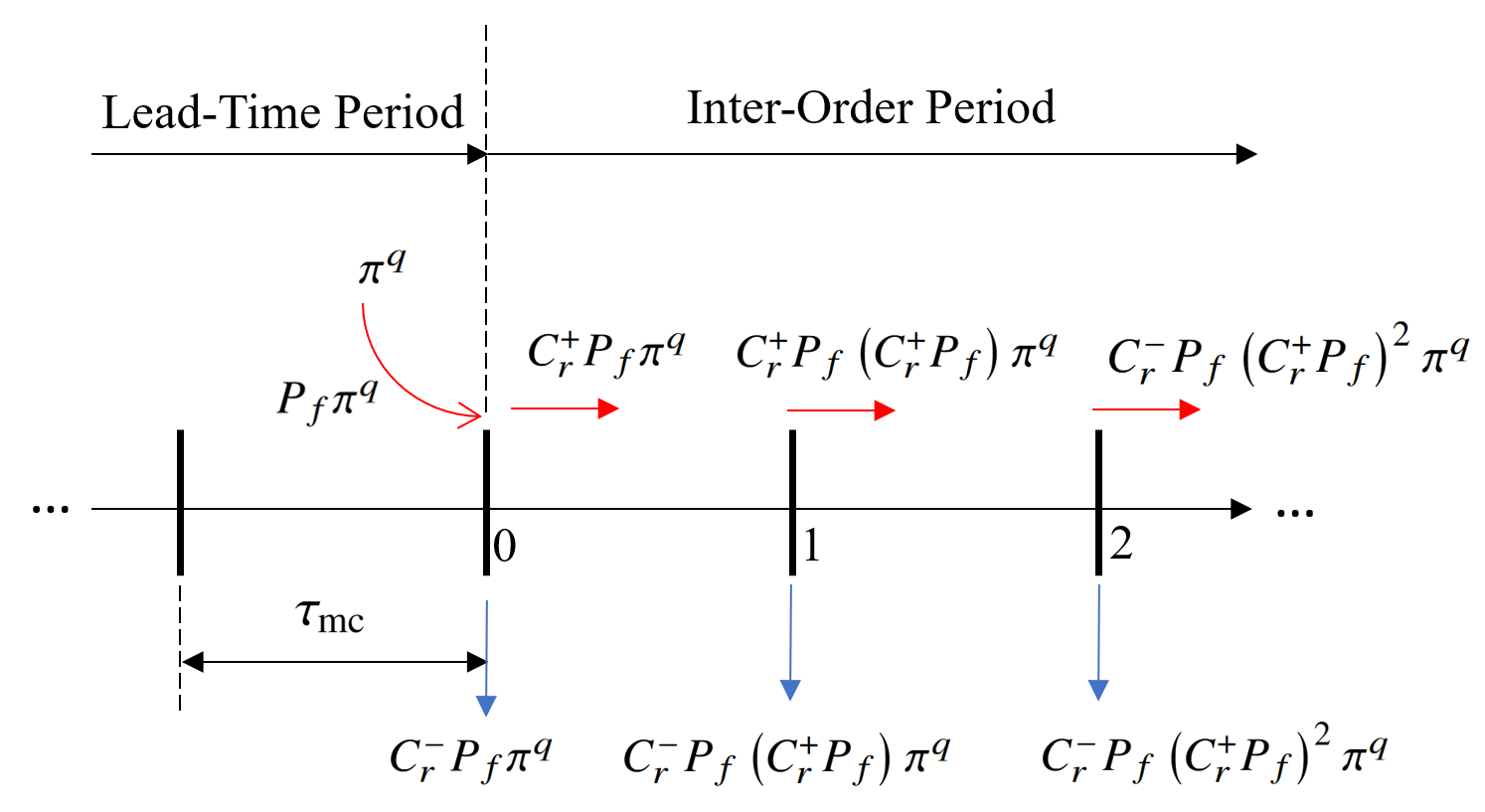}
    \caption{Transition diagram during IO period}
    \label{fig:direct_io_tran}
\end{figure}

This subsection derives the transition equation from $\pi^{q}$ to $\pi^{r}$. Assume that the reorder has just arrived before time step 0, as illustrated in Fig.~\ref{fig:direct_io_tran}. We now enumerate the possible state transitions following the receipt of $q$ replenishment units:
\begin{itemize}
    \item Distribution at time step 0, reorder is triggered: $C_r^- P_f \pi^{q}$
    \item Distribution at time step 1, reorder is triggered: $C_r^- P_f \left(C_r^+ P_f\right) \pi^{q}$
    \item Distribution at time step $j$, reorder is triggered: $C_r^- P_f \left( C_r^+ P_f \right)^j \pi^{q}$
\end{itemize}

The average distribution at the moment a reorder is eventually made is given by:
\begin{equation} \label{eq:piq_to_pir}
\begin{aligned}
    \pi^{r} &= \sum_{j=0}^\infty C_{r}^- P_f \left( C_r^+ P_f \right)^j \pi^{q} 
    = C_r^- P_f \left(I - C_r^+ P_f \right)^{-1} \pi^q
\end{aligned}
\end{equation}
Each of these events is mutually exclusive (i.e., a reorder cannot occur at multiple time steps simultaneously), and as \( j \to \infty \), the union of all events becomes collectively exhaustive. Since a reorder is guaranteed to occur eventually, the resulting expression requires no normalization. This formulation directly follows from the law of total expectation.

\subsection{Transition Equation from Reorder to Delivery}
\begin{figure}[!ht]
    \centering
    \includegraphics[width=.5\textwidth]{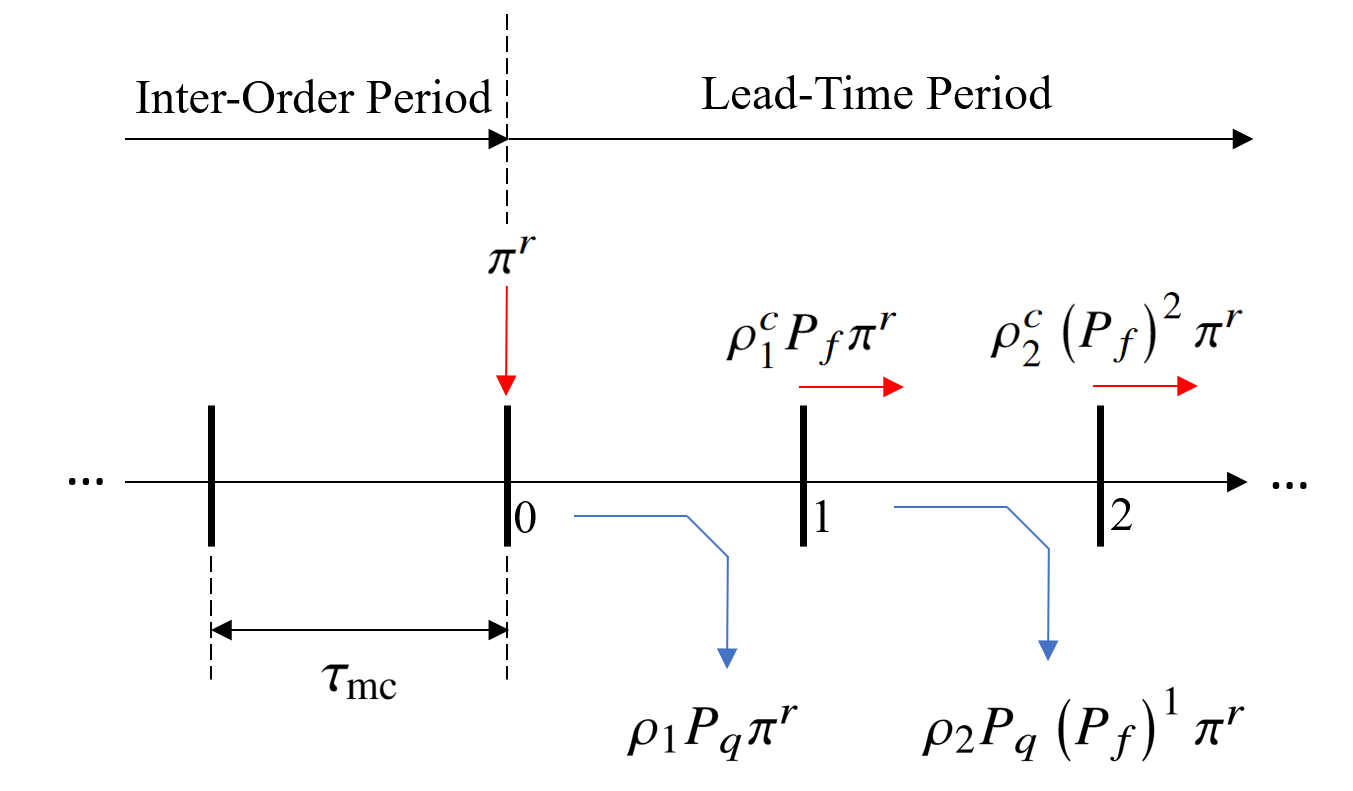}
    \caption{Timeline with the considered lead-time distribution}
    \label{fig:direct_lt_tran}
\end{figure}

This subsection derives the transition equation from $\pi^{r}$ to $\pi^{q}$. Consider the scenario where a reorder is placed at time step 0 with the expected state distribution $\pi^r$, as illustrated in Fig.~\ref{fig:direct_lt_tran}. Enumerating the all possible transitions are listed as: 
\begin{itemize}
    \item Distribution when replenishment arrives at time step 1: $\rho_1 P_q \pi^{r}$
    \item Distribution when replenishment arrives at time step 2: $\rho_2 P_q \left(P_f\right)^1 \pi^{r}$
    \item Distribution when replenishment arrives at time step $j$: $\rho_j P_q \left(P_f\right)^{j-1} \pi^{r}$
\end{itemize}

As in previous derivations, each event is mutually exclusive and collectively exhaustive. Therefore, the expected state distribution immediately after the replenishment, denoted by $\pi^q$, is computed as:
\begin{equation} \label{eq:direct_r2q}
   \pi^{q} = \sum_{j=0}^\infty \rho_{j+1} P_q \left(P_f\right)^{j} \pi^{r}
\end{equation}
For a general lead-time model, an approximated summation must typically be used. However, for the assumed lead-time distribution in Eq.~\eqref{eq:rho_lv}, an analytical expression for Eq.~\eqref{eq:direct_r2q} can be derived, and the result is given in Eq.~\eqref{eq:direct_r2q_anal}.

\subsection{Compute Distribution during Inter-Order (IO) Period}
This subsection derives the expression for $\pi^{\text{io}}$ in terms of $\pi^\text{q}$. Referring to Fig.~\ref{fig:direct_io_tran}, we enumerate the possible system states during the IO period as follows:
\begin{itemize}
    \item Distribution at time step 0, given that a reorder is not triggered: $C_r^+ P_f \pi^{q}$
    \item Distribution at time step 1, given that a reorder is not triggered: $C_r^+ P_f \left(C_r^+ P_f\right) \pi^{q}$
    \item Distribution at time step $j$, given that a reorder is not triggered: $C_r^+ P_f \left(C_r^+P_f \right)^j \pi^{q}$
\end{itemize}

The average state distribution during the IO period is then given by:
\begin{equation} \label{eq:pi_io_p_1}
\pi^{\text{io}} = \frac{1}{k_{\text{io}}} \sum_{j=0}^{\infty} C_r^+ P_f \left( C_r^+ P_f\right)^j \pi^{q}
= \frac{1}{k_{\text{io}}} C_r^+ P_f \left( I - C_r^+ P_f \right)^{-1} \pi^{q},
\end{equation}
where \( k_{\text{io}} \) is a normalization constant that ensures a valid probability distribution. It also represents the expected length of the IO period in units of \( \tau_\text{mc} \), yielding the time-converted interval as \( \tau_{\text{io}} = k_{\text{io}} \tau_\text{mc} \).

\subsection{Distribution during Lead-Time(LT) Period} 
This subsection derives the expression for $\pi^{\text{lt}}$ in terms of $\pi^{r}$. Referring to Fig.~\ref{fig:direct_lt_tran}, we enumerate the possible system states during the LT period as follows:
\begin{itemize}
    \item Distribution at time step 0, immediately after the order is placed: $\pi^{r}$
    \item Distribution at time step 1, given that replenishment has not arrived: $(1 - \rho_1)P_f\pi^{r}$
    \item Distribution at time step $j$, given that replenishment has not arrived: $\left(1 - \sum_{i=1}^{j}\rho_i\right)\left(P_f\right)^j\pi^{r}$
\end{itemize}
Here, the probability that replenishment has not arrived by the $j^\text{th}$ time step is defined as
\begin{equation}
    \rho^c_j = 1 - \sum_{i=1}^j \rho_i,\quad j=0,1,\dots,
\end{equation}
with $\rho_0^c = 1$. Thus, the weighted distribution at time step $j$ after the reorder, conditional on non-arrival, is $\rho^c_j \left(P_f\right)^j \pi^{{r}}$.

The average distribution during the LT period is then computed as
\begin{equation} \label{eq:pi_lt_p1}
    \pi^{\text{lt}} = \frac{1}{k_{\text{lt}}} \sum_{j=0}^\infty \rho^c_{j} \left(P_f\right)^j \pi^{r},
\end{equation}
where $k_{\text{lt}}$ is the normalization constant, which also represents the expected duration of the LT period in time steps. As before, we can derive the analytical expression of Eq.\eqref{eq:pi_lt_p1} based on the assumed lead-time distribution in Eq.\eqref{eq:rho_lv}, and the resulting expression is given in Eq.~\eqref{eq:pi_lt_anal}.

\subsection{Distribution during Every Replenishment Cycle}
Finally, since we have derived the state distributions during the IO and LT periods, along with their respective durations, the average state distribution in the constellation orbit over a complete replenishment cycle under the direct resupply policy is given by:
\begin{equation} \label{eq:pi_rc}
\pi^{\text{rc}} = \frac{k_{\text{io}}}{k_{\text{io}} + k_{\text{lt}}}\pi^{\text{io}} + \frac{k_{\text{lt}}}{k_{\text{io}} + k_{\text{lt}}}\pi^{\text{lt}},
\end{equation}
and the corresponding average duration of one replenishment cycle in the parking orbit is:
\begin{equation}
    \tau_{\text{rc}} = \tau_{\text{io}} + \tau_{\text{lt}}.
\end{equation}

\subsection{Flow of Direct Strategy Analysis}
To obtain the final result $\pi^\text{rc}$, a series of steps must be followed, as summarized in Table~\ref{alg:cap}. Unlike the indirect strategy analysis in \cite{han2025indirect}, the direct strategy does not require any iterative procedure, making the overall analysis significantly faster.

\begin{algorithm}
\caption{Analysis of Direct Strategy}\label{alg:cap}
\begin{algorithmic}
\Require Constellation Configuration, Probability Model
\State $P_{f} \gets$ Eq.~\eqref{eq:fail_matrix_sim} 
\State $C_{r_\text{p}}^-, C_{r_\text{p}}^+ \gets$ Eq.~\eqref{eq:Cr_matrix}
\State $P_{q} \gets$ Eq.~\eqref{eq:resupply_trans_matrix} 
\State $\pi^{q_\text{c}}$, $\pi^{r_\text{c}} \gets$ Eq.~\eqref{eq:piq_to_pir} and Eq.~\eqref{eq:direct_r2q}     
\State $\pi^\text{io}  \gets$ Eq.~\eqref{eq:pi_io_p_1}
\State $\pi^\text{lt}  \gets$ Eq.~\eqref{eq:pi_lt_p1}
\State $\pi^\text{rc}  \gets$ Eq.~\eqref{eq:pi_rc}
\end{algorithmic}
\end{algorithm}

\section{Performance Evaluation of Spare Management Policy} \label{sec4}
With the stationary solution \(\pi^{\mathrm{rc}}\) in hand, we can evaluate general performance metrics. The two most common metrics are operational cost and resilience, which typically trade off against each other.

\subsection{Cost Model of Direct Strategy}
The total expected operating cost per unit time, \(C_\text{total}\), is given by
\begin{equation}
    C_\text{total} = C_\text{build} + C_\text{hold} + C_\text{launch}, 
\end{equation}
where \(C_\text{build}\) is the expected manufacturing cost of spares per unit time,  
\(C_\text{hold}\) is the expected holding cost per unit time, and  
\(C_\text{launch}\) is the expected launch cost per unit time.

First, the manufacturing cost is defined as:
\begin{equation}
    C_\text{build} = \frac{1}{\tau_{\text{rc}}} c_\text{build}\  N_{\text{orbits}} \ q,
\end{equation}
where \(c_\text{build}\) is the manufacturing cost per spare satellite. This expression reflects that \(q\) spares are launched to each of the \(N_{\text{orbits}}\) orbits every \(\tau_{\text{rc}}\), and thus the same number of spares must be manufactured during each replenishment cycle.

The holding cost represents the penalty for maintaining an excessive number of spares in orbit. It accounts for station-keeping, depreciation, and failure risk. It is modeled as:
\begin{equation}
    C_\text{hold} = c_{\text{hold}}  N_{\text{orbit}} \sum_{i=\bar{N}_\text{sat}+1}^{N_{\text{sat}}} \left( i - \bar{N}_\text{sat} \right) 
    \pi^{\text{rc}}(X = i),
\end{equation}
where \(c_{\text{hold}}\) is the holding cost per spare satellite per unit time in constellation orbits. The summation computes the expected number of spare satellites.

Lastly, the expected launch cost is modeled under two scenarios, depending on whether rideshare opportunities are available:
\begin{equation} 
C_\text{launch} = 
\begin{cases} 
\dfrac{N_{\text{orbit}}}{\tau_{\text{rc}}} \min \bigl\{\, c_\text{lv,unit}\, m_\text{total},\; c_\text{lv,full} \,\bigr\}, & \text{if rideshare available}, \\
\dfrac{N_{\text{orbit}}}{\tau_{\text{rc}}}\, c_\text{lv,full}, & \text{if rideshare unavailable}. 
\end{cases} 
\end{equation}
Here, $c_\text{lv,unit}$ is the launch cost per unit mass to LEO, $c_\text{lv,full}$ is the discounted cost of reserving the full vehicle, and $m_\text{total} = m_\text{sat} q$ is the total payload mass of $q$ satellites with individual mass $m_\text{sat}$. The first case reflects the trade-off between per-unit and full-contract pricing when rideshare missions to the target orbit are available. The second case assumes that rideshare is not offered to the desired orbit, so the operator must always purchase the full vehicle.

\subsection{Resilience Model of Direct Resupply Strategy}
The proper resilience metric should capture both agility (how quickly the system recovers to nominal capacity) and robustness (the depth of performance degradation while below nominal) \cite{dod_resilience_2011, Ron_resilience}. For consistency, we use the same metric defined in \cite{han2025indirect}. In the discrete-time Markov model, resilience is measured by the expected shortage \(S\):
\begin{equation}
    S = \sum_{i=0}^{\bar{N}_\text{sat}} (\bar{N}_{\text{sat}} - i)\cdot \pi^{\text{rc}} (X=i),
\end{equation}
which weighs the deficit \(\bar{N}_{\text{sat}} - i\) by the fraction of time the system spends in state \(i\), as given by \(\pi^{\text{rc}} (X=i)\).

\subsection{Optimization Problem for Direct Resupply Strategy}
There are multiple ways to formulate the optimization problem, but here we focus on minimizing the total operating cost of the spare policy while enforcing resilience and launch vehicle constraints. The problem is formulated as:
\begin{equation} \label{eq:opt_formulation}
\begin{aligned}
    \min_{x}\quad & C_\text{total}\\
    \text{s.t.}\quad & g_1 = S - \varepsilon \leq 0, \\
    & g_2 = m_\text{total} - m_\text{payload} \leq 0, \\
    & q, r \in \mathbb{Z}^+,
\end{aligned}
\end{equation}
where $x = (q, r)$, $\varepsilon$ is a user-defined threshold for the acceptable shortage level, $g_1$ enforces the resilience constraint, and $g_2$ ensures that the total launch mass does not exceed the payload capacity $m_\text{payload}$ of the launch vehicle. Solving this problem provides an estimate of the cost required to maintain the satellite constellation under the direct strategy.

\section{Numerical Validation of the Analysis Method} \label{sec5}
\subsection{Numerical Validation Set-up}
The analytical model developed above enables efficient evaluation of spare policies even for mega-scale constellations, but it must be validated before being applied to other use cases. To validate the proposed method, we follow the approach introduced in \cite{jakob2019optimal}.

While one could directly compare the histogram of simulated stock levels with \(\pi^{\mathrm{rc}}\), summarizing the differences in a single metric is challenging. Therefore, we adopt the same metrics used in \cite{han2025indirect}: the mean stock level and the expected shortage in the constellation orbit. The mean stock level is computed as
\begin{equation}
    M = \sum_{i=0}^{N_{\text{sat}}} i\cdot \pi^{\text{rc}} (X=i).
\end{equation}

We construct 100 unique test cases using Latin hypercube sampling over the parameter ranges in Table~\ref{tab:trade_space_lhs}, with fixed parameters provided in Table~\ref{tab:fixed_sim_para}. For each test case, a 20-year simulation is run 1000 times, and the results are averaged to obtain the final statistics.

\subsection{Numerical Validation Result}
\begin{figure}[!ht]
    \centering
    \includegraphics[width=.45\textwidth]{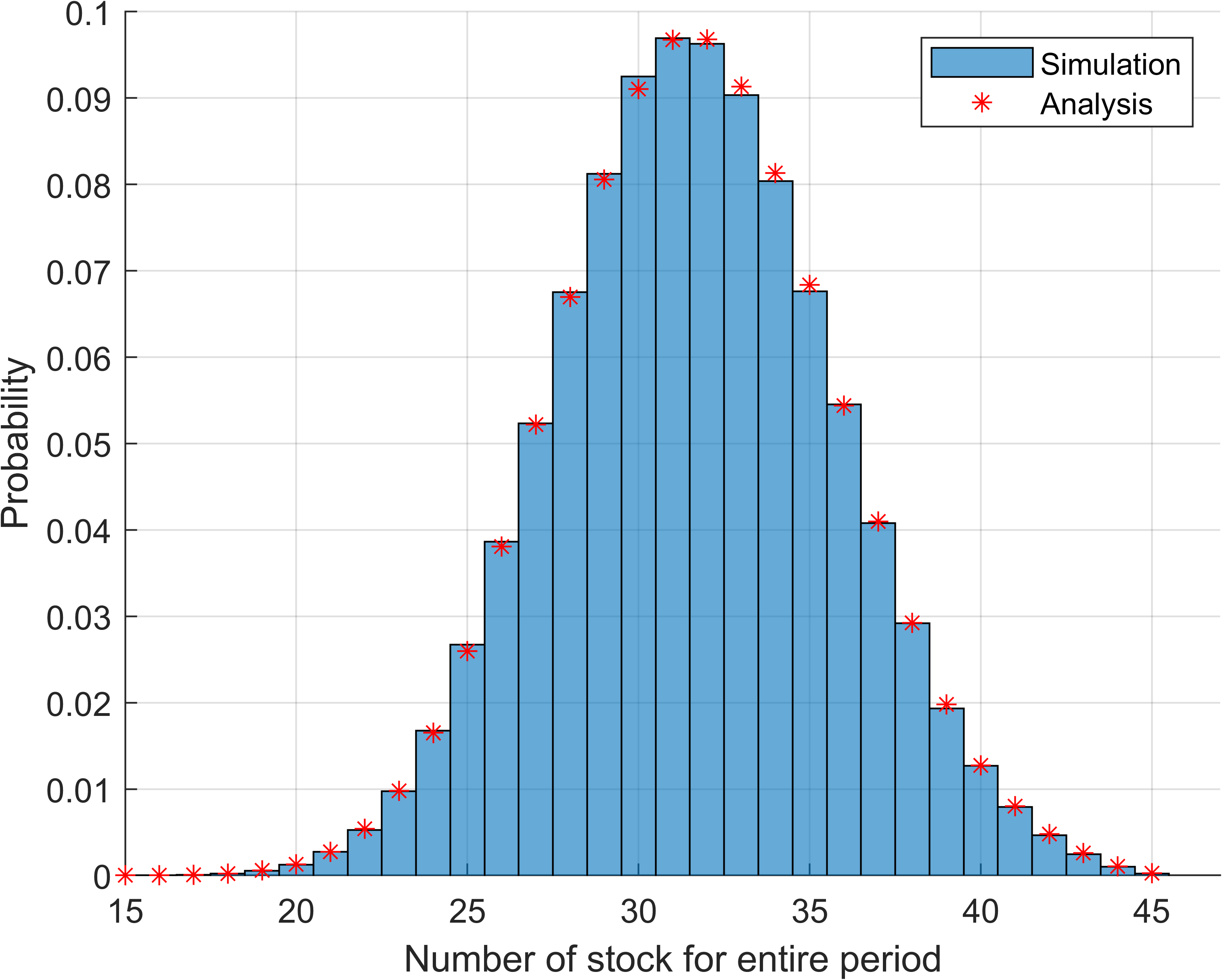}
    \caption{ $\pi^{\text{rc}}$ for the representative case of large error}
    \label{fig:worst_case_result}
\end{figure}

The test results are summarized in Table~\ref{tab:error_list}. Relative error is used to quantify discrepancies between the analytical model and the simulation. The errors arise primarily from Monte Carlo noise and time step discretization. Nevertheless, the maximum relative error across all metrics is below 1\%. The larger error observed for $S$ compared with $M$ occurs when $S$ is close to zero, which amplifies the relative error. Even in the worst cases, direct comparisons of the probability distributions show that the proposed method accurately captures system behavior. A representative test case with near-maximum error across all four metrics is illustrated in Fig.~\ref{fig:worst_case_result} ($\lambda = 0.35,\ q=2,\ r = 43, \mu_\text{lv} = 10, \tau_\text{lv} = 60$).

Finally, in terms of computational efficiency, the full simulation required several hours to complete, whereas the proposed analytical method computed each test case in under a millisecond. This confirms that the method is both accurate and computationally efficient, making it suitable for use as the inner loop of an optimization process.

\begin{table}[hbt!]
\centering
\caption{Fixed simulation parameters}
\label{tab:fixed_sim_para}
\begin{tabular}{lc c c c}
\hline \hline
Parameter & Notation & Value & Unit \\
\hline

Markov time step
  & $\tau_\text{mc}$ 
  & $0.5$ 
  & days \\

Number of constellation orbits
  & $N_{\text{orbit}_\text{c}}$ 
  & $40$ 
  & orbits \\

Nominal satellites per plane
  & $\bar{N}_\text{sat}$ 
  & $40$ 
  & satellites \\

\hline
\hline
\end{tabular}
\end{table}

\begin{table}[hbt!]
\centering
\caption{Bound of sampled simulation parameters}
\label{tab:trade_space_lhs}
\begin{tabular}{lc c c c}
\hline \hline
Parameter & Notation & Bounds & Unit \\
\hline

Satellite failure rate
  & $\lambda_{\text{sat}}$ 
  & $\left[ 0.001,\  0.5\right]$ 
  & failures/satellite/year \\

Launch order processing time 
  & $\tau_{\text{lv}}$ 
  & $\left[ 0,\  60\right]$ 
  & days \\

Mean exponential launch lead time
  & $\mu_{\text{lv}}$ 
  & $\left[ 5,\  60\right]$ 
  & days \\

Order size for in-plane spares
  & $q$ 
  & $\left[ 1,\  10\right]$ 
  & satellites \\

Reorder point for in-plane spares
  & $r$ 
  & $\left[\bar{N}_\text{sat}- 5,\  \bar{N}_\text{sat}+5\right]$ 
  & batches \\  

\hline
\hline
\end{tabular}
\end{table}

\begin{table}[hbt!]
\centering
\caption{Error between proposed method and simulation results}
\label{tab:error_list}
\begin{tabular}{lc c c }
\hline \hline
Parameter & Mean & P95\\
\hline
Relative error of $M$ 
  & $0.026 \ \%$ 
  & $0.097 \ \%$ \\

Relative error of $S$ 
  & $0.221 \ \%$ 
  & $0.802 \ \%$ \\

\hline
\hline
\end{tabular}
\end{table}

\section{Optimization of Spare Management Policy} \label{sec6}
A key application of the proposed analysis method is design optimization. Based on stakeholder interests, an optimization problem can be formulated to guide early-phase decisions—such as how many spares to prepare per cycle or the resulting monthly cost. In this section, we optimize the direct strategy using the formulation in Eq.~\eqref{eq:opt_formulation}, with real-world parameters.

For the direct strategy, we assume the use of Rocket Lab’s Electron as the launch vehicle, with launch cost data referenced from \cite{SpaceInsider_cost}. Although the nominal lead time for Electron can be as short as two days \cite{ainvest_rocketlab_2025}, we adopt a more conservative estimate of \((\mu_\text{lv}, \tau_\text{lv}) = (10, 10)\) days to account for contractual and logistical delays beyond vehicle readiness. Note that the LV unit cost is chosen so that launching the full payload at the per-kilogram rate incurs a 20\% premium compared to reserving the entire vehicle, i.e., $c_\text{lv,unit} \cdot m_\text{payload} = 1.2 \cdot c_\text{lv,full}$. All remaining parameters are listed in Tables~\ref{tab:fixed_sim_para} and \ref{tab:opt_para}.

\begin{table}[hbt!]
\centering
\caption{Parameters for the optimization}
\label{tab:opt_para}
\begin{tabular}{lc c c c}
\hline \hline
Parameter & Notation & Value & Unit \\
\hline

Satellite manufacturing cost
  & $c_{\text{build}}$ 
  & $0.5$ 
  & M\$/satellite \\

In-orbit spares annual holding cost
  & $c_{\text{hold}}$ 
  & $0.5$ 
  & M\$/satellite/year \\

Launch cost per unit mass (Electron)
  & $c_{\text{lv,unit}}$ 
  & $30000$ 
  & \$/kg \\

Discounted cost for full contract (Electron)
  & $c_{\text{lv,full}}$ 
  & $7.5$ 
  & M\$ \\

Payload launch maximum capacity (Electron)
  & $m_{\text{payload}}$ 
  & $300$ 
  & kg \\

Mass of satellite
  & $m_\text{sat}$ 
  & $150$ 
  & kg \\

\hline
\hline
\end{tabular}
\end{table}

\subsection{Baseline Scenario}
As a baseline scenario, we consider a moderate failure case with \(\lambda_\text{sat} = 0.05\) and set the shortage threshold to \(\varepsilon = 0.25\). Due to the small design space, enumerating all possible cost values and constraint feasibility can be done, and the optimal solution found for this scenario is \(x^\ast = (q^\ast, r^\ast) = (2, 39)\). The total cost, cost breakdown, and constraint feasibility are summarized in Table~\ref{tab:base_scn}. It turns out that full contract minimizes the cost even when rideshare is available.

\begin{table}[hbt!]
\caption{Results summary of representative scenarios \label{tab:base_scn}}
\centering
\begin{tabular}{l c c ccc c c}
\hline \hline
    \multirow{2}{*}{Rideshare} &
    \multirow{2}{*}{$C_\text{total}$ [M\$/day]} && \multicolumn{3}{c}{Detailed Costs [M\$/day]} && \multicolumn{1}{c}{Constraints} \\
    \cline{4-6} \cline{8-8} 
  & && $C_\text{build}$ & $C_\text{hold}$ &$ C_\text{launch}$  &&  $S$  \\
\hline\hline
 Allowed & 0.9547  && 0.1094 & 0.0246 & 0.8207  && 0.0591 \\
 Not Allowed & 0.9547  && 0.1094 & 0.0246 & 0.8207  && 0.0591 \\
\hline \hline
\end{tabular}
\end{table}

\subsection{Sensitivity to Failure Rate}
Satellite failure rates can vary over time and depend on the scale of the constellation. To assess robustness, we evaluate a wide range of failure rates from 0.001 to 0.5 failures per year under the assumption that rideshare is available.

In this experiment, we vary only the failure rate while keeping all other parameters fixed as in the baseline scenario. Figure~\ref{fig:cost_sens} shows how the total cost and its components change with the failure rate, and Fig.~\ref{fig:para_sens} represents how the optimal solution evolves.

The results show that contracting the full payload capacity is optimal in all cases except the region where $\lambda_\text{sat} \in [0.001,\ 0.008]$, where failures are nearly negligible. In this low-failure regime, rideshare can reduce the total cost.
As the failure rate increases, the optimal reorder point increases (from 39 to 42), as expected, to maintain the desired resilience level.

\begin{table}[hbt!]
\caption{Results summary of representative scenarios \label{tab:base_scn}}
\centering
\begin{tabular}{l c c ccc c c}
\hline \hline
    \multirow{2}{*}{Rideshare} &
    \multirow{2}{*}{$C_\text{total}$ [M\$/day]} && \multicolumn{3}{c}{Detailed Costs [M\$/day]} && \multicolumn{1}{c}{Constraints} \\
    \cline{4-6} \cline{8-8} 
  & && $C_\text{build}$ & $C_\text{hold}$ &$ C_\text{launch}$  &&  $S$  \\
\hline\hline
 Allowed & 0.9547  && 0.1094 & 0.0246 & 0.8207  && 0.0591 \\
 Not Allowed & 0.9547  && 0.1094 & 0.0246 & 0.8207  && 0.0591 \\
\hline \hline
\end{tabular}
\end{table}

\subsection{Sensitivity to Failure Rate}
Satellite failure rates can vary over time and depend on the system scale. Therefore, we evaluate a wide range of failure rates from 0.001 to 0.5 failures per year to assess how the optimal solution changes with assumption of rideshare available.

\begin{figure}[!ht]
    \centering
    \includegraphics[width=.45\textwidth]{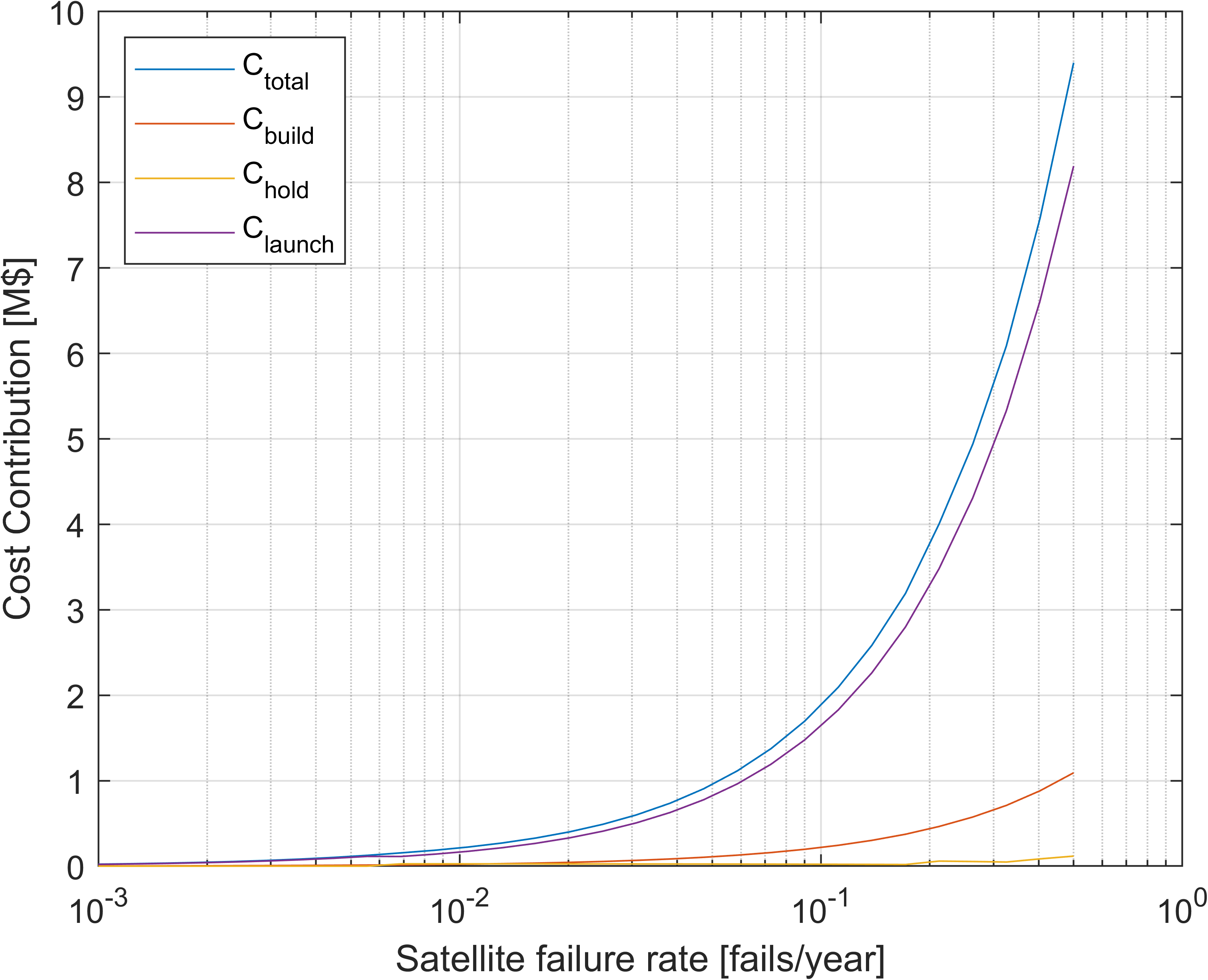}
    \caption{Cost contribution of direct strategy with respect various failure rates}
    \label{fig:cost_sens}
\end{figure}

\begin{figure}[!ht]
    \centering
    \includegraphics[width=.45\textwidth]{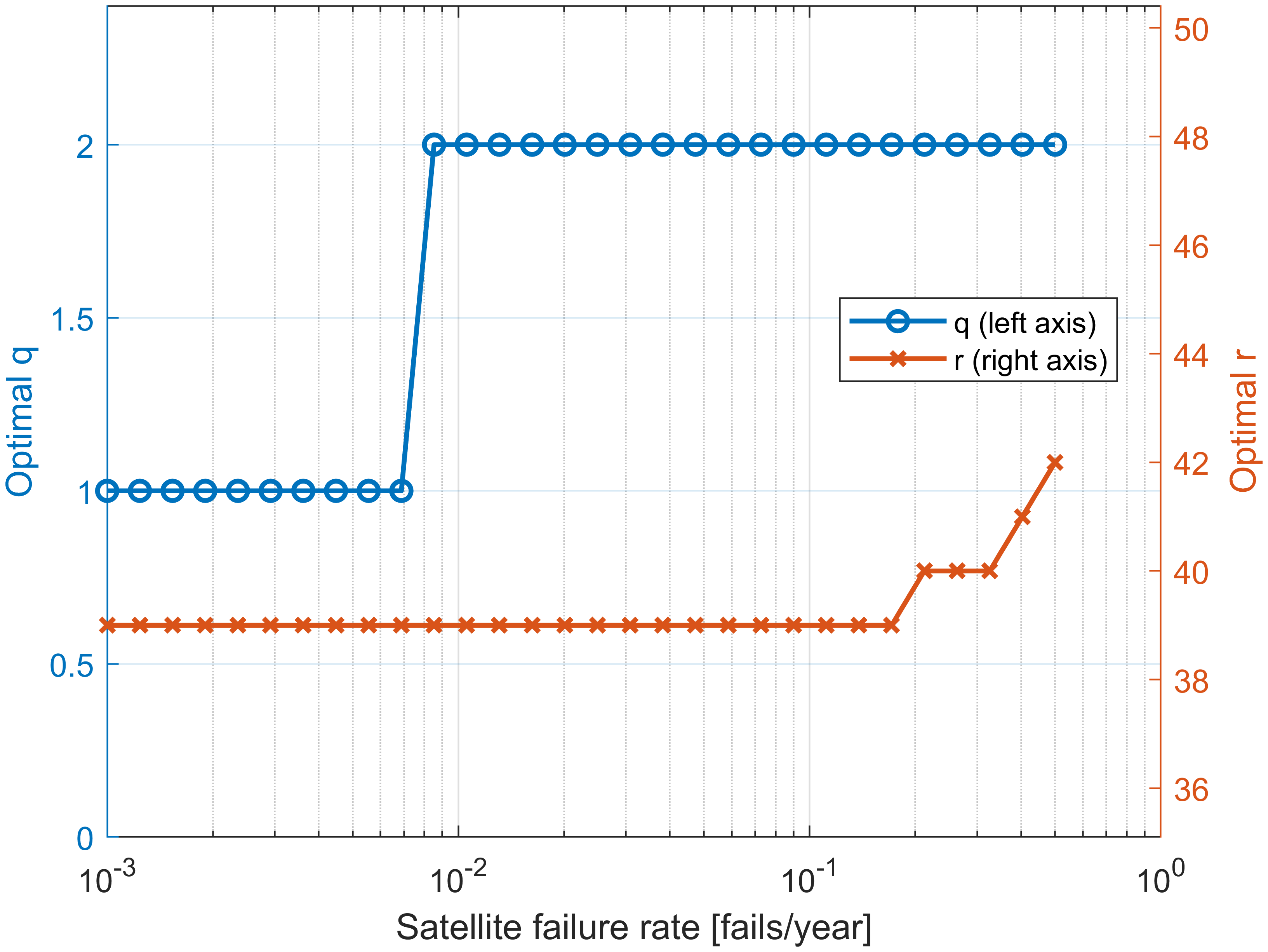}
    \caption{Optimal design of direct strategy with respect various failure rates}
    \label{fig:para_sens}
\end{figure}

In this experiment, we vary only the failure rate while keeping all other parameters fixed as in the baseline scenario. Figure~\ref{fig:cost_sens} shows how the total cost and its components change with the failure rate, and Fig.~\ref{fig:para_sens} represents how the optimal solution evolves.

The results show that contracting the full payload capacity is optimal in all cases except the region where $\lambda_\text{sat} \in [0.001,\ 0.008]$, where failures are nearly negligible. Therefore, for that reasons, having rideshare could reduce the total cost.
As the failure rate increases, the optimal reorder point increases (from 39 to 42), as expected, to maintain the desired resilience level.


\section{Conclusion} \label{sec7}
In this paper, we developed a Markov chain–based framework for the detailed analysis and design of a direct spare management policy for large-scale constellations. We modeled each in-plane orbit as an \((r, q)\) system and derived the expressions for both cost and resilience metrics. Building on this fast and accurate analysis, we formulated and solved an optimization problem to minimize the total operating cost subject to resilience constraints. The optimized result for the direct strategy is then used for comparison with the indirect strategy in a companion study~\cite{han2025indirect}. Finally, the proposed framework can be extended to support other constellation configurations (e.g., heterogeneous agents) and alternative replenishment policies, which we will explore in future work.

\section*{Appendix}
In this appendix, we will derived the analytical expression using the explicit equation of exponential lead time distribution.

\subsection*{Analytic Expression for Eq.~\eqref{eq:direct_r2q}}
Substituting Eq.~\eqref{eq:rho_lv} into Eq.~\eqref{eq:direct_r2q} gives
\begin{equation} \label{eq:direct_r2q_anal}
\begin{aligned}
    \pi^{q} &= \sum_{j=m}^\infty \alpha^{j-m} \left( 1 - \alpha\right)  P_q \left(P_f\right)^j \pi^{r} \\
    &= \left( 1 - \alpha\right)P_q \left(P_f\right)^{m} \left( I + \alpha P_f + \alpha^2 \left(P_f\right)^2 + \cdots \right) \\
    &= \left( 1 - \alpha\right) P_q \left(P_f\right)^{m}\left(I-\alpha P_f \right)^{-1}\pi^r
\end{aligned}
\end{equation}
Here, $\alpha = e^{-\tau_\text{mc}/{\mu_\text{lv}}}$, and $m = \lceil \tau_\text{lv}/\tau_\text{mc} \rceil$, where $\lceil \cdot \rceil$ denotes the ceiling operator. Note that $m$) represents the minimum number of discrete time steps required to complete the fixed portion of the lead time.

\subsection*{Analytic Expression for Eq.~\eqref{eq:pi_lt_p1}}
As before, substituting Eq.~\eqref{eq:rho_lv} into Eq.~\eqref{eq:pi_lt_p1} gives
\begin{equation} \label{eq:pi_lt_anal}
\begin{aligned}
    \pi^{\text{lt}} &= \frac{1}{k_{\text{lt}}}
    \left(  \sum_{i=0}^m \left(P_f\right)^i + \sum_{i=m}^\infty \rho^c_{i-m} \left(P_f\right)^i \right) \pi^{r} \\
    &= \frac{1}{k_{\text{lt}}} \left(\sum_{i=0}^m \left(P_f\right)^i  + \alpha \left(P_f\right)^{m+1}\left( I - \alpha P_f\right)^{-1} \right)\pi^r. 
\end{aligned}
\end{equation}
Note that approximation error may occur when $\tau_\text{lv}$ is not an integer multiple of $\tau_\text{mc}$. While the equation can be further refined to correct this approximation, one can also avoid the issue by selecting $\tau_\text{mc}$ as an integer divisor of $\tau_\text{lv}$, or by choosing $\tau_\text{mc}$ small enough to make the approximation error negligible.

\section*{Funding Sources}
This research was supported by the Advanced Technology R\&D Center at Mitsubishi Electric Corporation.


\bibliography{references}

\end{document}